\documentclass[sigconf,9pt]{acmart}

\usepackage[english]{babel}
\usepackage[T1]{fontenc}
\usepackage[utf8]{inputenc}
\usepackage{cleveref}
\usepackage{tikz}
\usetikzlibrary{positioning,shapes,arrows,decorations.text,automata,arrows.meta}
\usepackage{listings}
\usepackage[separate-uncertainty=true,multi-part-units=single]{siunitx}
\usepackage{balance}

\usepackage[absolute,overlay]{textpos}
\newif\ifauthorversion
\authorversiontrue

\setcopyright{acmcopyright}



\acmDOI{10.1145/3229591.3229593}

\acmISBN{978-1-4503-5908-5/18/08}

\acmConference[NetCompute'18]{ACM SIGCOMM 2018 Morning Workshop on In-Network Computing }{August 20, 2018}{Budapest, Hungary}
\acmBooktitle{NetCompute'18: ACM SIGCOMM 2018 Morning Workshop on In-Network Computing , August 20, 2018, Budapest, Hungary}
\acmYear{2018}
\copyrightyear{2018}

\acmPrice{15.00}

\begin{document}
\title{P4CEP: Towards In-Network Complex Event Processing}
\renewcommand{\shorttitle}{P4CEP: Towards In-Network Complex Event Processing }

 \author{Thomas Kohler, Ruben Mayer, Frank Dürr, Marius Maaß, Sukanya Bhowmik, and Kurt Rothermel}
 \orcid{0000-0002-2990-2257}
 \affiliation{%
   \institution{Institute of Parallel and Distributed Systems (IPVS), University of Stuttgart}
   \streetaddress{Universitätsstr. 38}
   \city{Stuttgart} 
   \state{Germany} 
   \postcode{70569}
 }
 \email{{firstname}.{lastname}@ipvs.uni-stuttgart.de}

\renewcommand{\shortauthors}{Kohler et al.}

\begin{abstract}
%
%
%
In-network computing using programmable networking hardware is a strong trend in networking that promises to reduce latency and consumption of server resources through offloading to network elements (programmable switches and smart NICs). In particular, the data plane programming language P4 together with powerful P4 networking hardware has spawned projects offloading services into the network, e.g., consensus services or caching services. In this paper, we present a novel case for in-network computing, namely, Complex Event Processing (CEP). CEP processes streams of basic events, e.g., stemming from networked sensors, into meaningful complex events. Traditionally, CEP processing has been performed on servers or overlay networks. However, we argue in this paper that CEP is a good candidate for in-network computing along the communication path avoiding detouring streams to distant servers to minimize communication latency while also exploiting processing capabilities of novel networking hardware. We show that it is feasible to express CEP operations in P4 and also present a tool to compile CEP operations, formulated in our P4CEP rule specification language, to P4 code. Moreover, we identify challenges and problems that we have encountered to show future research directions for implementing full-fledged in-network CEP systems.

\end{abstract}

\maketitle

\ifauthorversion
\begin{textblock*}{\paperwidth}(0mm,26cm)
\centering
© ACM 2018. This is the author's version of the work. It is posted here for your personal use. Not for redistribution.\\
The definitive Version of Record was published in NetCompute'18, https://dx.doi.org/10.1145/3229591.3229593.
\end{textblock*}
\fi

%
%
%
\section{Introduction}
Recent developments in Software-defined Networking (SDN) have given rise to a new evolutionary step of network programmability.
Exploiting data plane programming for offloading of application functionality from end-systems to programmable network elements while leveraging the performance of specialized forwarding hardware, capable of processing packets at line-rate throughput in orders up to Terabits per second with low latency, is a recent trend in networking called \emph{in-network computing}.
In-network computing has so far been proposed to support various distributed applications, ranging from consensus \cite{dang_netpaxos:_2015}, over caching in distributed key-value stores \cite{liu_incbricks:_2017} and network diagnostics, to aggregation functions in data-centric processing including machine learning and graph analytics \cite{sapio_-network_2017}. It has been shown that in-network computing can yield significant performance improvements by increasing throughput, bandwidth-efficiency, or reducing latency.

In this work, we employ in-network computing to offload \emph{Complex Event Processing} (CEP), a representative of stateful processing from the domain of message-oriented middleware. Traditionally, CEP has been implemented as an overlay of software middleboxes (operators) inferring higher-level knowledge (complex events) by evaluating specific combinations of incoming information (basic events). Packets convey basic events, which are typically comprising structured low-dimensional data, such as sensor data, stock market values for high-frequency trading, or data of network management, such as intrusion-detection systems or anomaly detection.
In-network computing is best suited for processing of small data encapsulated in packet headers. Furthermore, CEP systems seek to reduce processing latency, e.g., in high-frequency trading, and to optimize bandwidth utilization. These goals are congruent to the performance gains of in-network computing.

We find \cite{kohler_infep_2017} that a large class of applications, including CEP, is based on the \emph{middlebox model}, where packets are processed on remote hardware appliances (middleboxes) or in virtualized environments on commodity server hardware (NFV). Although CPUs are cheap and allow for arbitrary packet processing in software, which has become remarkably fast, the middlebox model bears disadvantages: it increases network management complexity by the introduction of additional system components (remote, off-path entities) that can fail, and have to be managed (placement, dynamic configuration). By steering traffic through remote hardware, additional round trips are inherently inflicted, consequently increasing application latency, which is further exacerbated by \emph{service chaining}.
Thus, packets are ideally processed \emph{in-situ} at high-performance network elements that they naturally traverse, consequently combining forwarding and processing, which resembles the rationale of in-network computing.
Furthermore, the uniform interface of data plane programming, provided by the P4 language, greatly facilitates portability.

In previous work \cite{bhowmik_high_2017}, we have offloaded content-based filtering in publish/subscribe middleware systems from middleboxes (\emph{brokers}) to traditional SDN-enabled switches with a fixed pipeline. While this proved to be sufficient for \emph{filtering} messages at line-rate, we now leverage in-network computing for the in-situ \emph{processing} of complex events in the network, which is challenging due to current limitations of data plane programming for stateful packet processing.

Our contributions in this paper are as follows: we present P4CEP---our early work on an in-network implementation of CEP, including a proof-of-concept compiler from our P4CEP rule specification language to P4. We show that our design contains generic mechanisms, namely window-based aggregation functions (for reasoning over a window of events) and state-machine logic (for event detection), which are highly relevant and can be reused for the in-network implementation of other stateful packet processing applications.
We discuss requirements from the perspective of CEP applications and provide feedback on useful data plane programming aspects and experienced limitations.
We provide a preliminary evaluation of the performance properties of our implementation, which we deployed on programmable NIC hardware targets.
Furthermore, we lay out a roadmap to a distributed in-network CEP implementation, addressing replication and partitioning strategies as well as in-network pre-filtering.
The implementation of our prototype of P4CEP is available at: \url{https://goo.gl/MEdPvv} \cite{p4cep-release}.

%
%
%
\section{Background}
\label{sec:background}
In this section, we give a brief introduction to data plane programming and describe limitations imposed by P4 and existing hardware targets, followed by an introduction to Complex Event Processing.

\subsection{Data Plane Programming with P4}
\label{subsec:background:P4}
The paradigm of \emph{data plane programming} subsumes the combination of (1) a quasi-standardized, hardware-agnostic domain-specific language (P4) implementing a uniform interface for defining the forwarding behavior of (2) emerging reconfigurable data plane hardware.
Key elements of reconfigurable hardware are a parser defining header syntax and semantics and a match-action engine defining the semantics of processing. Both, parser and engine are software-definable, implementing a programmable multi-stage pipeline.
Due to space constraints, we omit the description of the P4-language, while referring to \cite{bosshart_p4}.

For P4CEP, we consider the following targets: hardware targets residing in end-systems, such as (1) the Netronome Agilio NIC (NFP framework)\cite{open-nfp}, (2) the NetFPGA platform, as well as (3) reconfigurable ASIC-based (RMT) or FPGA-based (Corsa) data center switches. Furthermore, we consider (4) software-switch implementations, such as the P4 reference switch implementation bmv2 and PISCES, and (5) extended Berkeley Packet Filter (eBPF), which provide fast in-kernel processing within end-systems

In general, hardware targets face inherent limitations \cite{sapio_-network_2017, jose_compiling_2015}.
(1)~The size of both SRAM and TCAM memory is limited, which imposes bounds on the number of tables, their entries, and other held state.
(2)~Hardware switches are designed to unconditionally guarantee line-rate throughput. This places an upper bound on processing latency in the order of tens of nanoseconds, consequently bounding the number and complexity of packet operations in each pipeline stage. Hence, P4 models the control flow as an imperative program that specifies the execution sequence through the pipeline as a DAG, which rules out loops and thus renders P4 Turing-incomplete.
(3)~Stateful packet processing on programmable switches has been shown to be challenging \cite{sivaraman_packet_2016}.
Unsynchronized, concurrent access can lead to inconsistency effects, such as lost updates, which pose a severe threat for the correctness of stateful packet processing algorithms. The support of atomic register operations is target-dependent and not mandated by P4. However, Netronome's NFP SDK provides a pre-processor pragma for global synchronization of register access.

While reconfigurable switching ASICs are primarily designed for networking tasks like forwarding, FPGAs are much more flexible as they allow for the implementation of custom logic in hardware. To be able to exploit the extended programmability of such targets, $\text{P4}_\text{16}$ includes the \texttt{extern} primitive, which provides an interface to functions that are not part of the P4 specification, such as checksum computation and cryptographic operations. They can also be used for synchronization of register access. For instance, the NFP framework allows referencing to external functions written in C and executed in a C-sandbox running on NFP's micro-engines. It natively supports efficient atomic arithmetic operations and has a built-in mutex and semaphore library. Although external functions are a very powerful concept, they break target-independence and possibly lead to unbounded processing latency.

\subsection{Complex Event Processing}
\label{subsec:background:CEP}
Complex Event Processing (CEP) is a paradigm to infer the occurrence of situations of interest from basic events \cite{cugola_tesla:_2010}. For instance, in the field of algorithmic trading, a situation of interest can be the detection of a leading market signal, whereas the basic event streams contain stock quotes of a stock exchange.
An example from the field of sensor fusion is detecting fire (the complex event) by reasoning over measurements of networked smoke and temperature sensors (basic events).
In doing so, a CEP system deploys a distributed \emph{operator graph} between event sources and sinks, where each operator detects a specific event pattern in its input streams and emits output events when instances of the corresponding event pattern have been detected. 

The pattern to be detected by a CEP operator is typically defined in an event specification language \cite{cugola_tesla:_2010} as a \emph{continuous query}. Such a query consists of a number of matching expressions, such as Sequence, AND, OR, NOT, etc., that specify the conditions under which a sequence of input events matches the query. Furthermore, a query can contain aggregation operations such as MAX, MIN, AVG, etc., that are known from stream processing systems \cite{Arasu:2006:CCQ:1146461.1146463}. In the fire detection example, these expressions are used to combine measurements of different sensor types (smoke, temperature) and allow the reasoning over their aggregated measurement values. Based on existing languages, we define a meaningful subset for in-network CEP, which we describe in detail in section \ref{subsec:P4CEP:language}.

CEP operators are often stateful, i.e., the processing of one event may influence the internal state of the operator, which in turn influences the processing of subsequent events. Usually, the state relevant to a CEP operator is limited by a sliding window \cite{Mayer:2017:SSC:3135974.3135983}. A sliding window restricts the infinite sequence of input events in an operator to a subsequence that can match the query. The extents of a sliding window are specified by a \emph{window policy}, which defines the size of a window and its slide, i.e., by how much the window moves from one window instance to the next. In the example, sliding windows enable reasoning over time-series of measurements, e.g., allowing to infer trends. For instance, a fire can be defined to be inferred, when the averages over the last $n$ measurements of the smoke and temperature sensors exceed a given threshold.

Thus, both pattern detection and sliding windows require holding state among the processing of incoming events. For an in-network implementation this consequently mandates stateful processing of packets (events), holding and processing per-packet state as well as inter-packet state. Required consistency semantics on reliability (lost events) and ordering (out-of-order events) in event processing may differ depending on the CEP application. We address consistency implications for P4CEP in section \ref{subsec:P4CEP:workflow-compiler}.

Typically, CEP operators are executed on end-systems. Typical performance figures show average processing latencies of about \SI{200}{\micro\second}, excluding the end-system's network stack latency, for detecting sequences of two states \cite{cugola_complex_2012}, which is the simplest form of stateful processing. In terms of throughput, highly parallel implementations of CEP operators on multi-core CPUs can reach up to 218,000 events/second for more complex patterns \cite{Mayer:2017:SSC:3135974.3135983}.

%
%
%
\section{P4CEP: Design \& Implementation}
\label{sec:p4cep}
In this section, we present our underlying system model, the P4CEP-compiler, and the P4CEP rule specification language along with an example illustrating our design.

\subsection{System Model}
\label{subsec:P4CEP:system-model}
\begin{figure}
	\centering
		\includegraphics[width=\columnwidth]{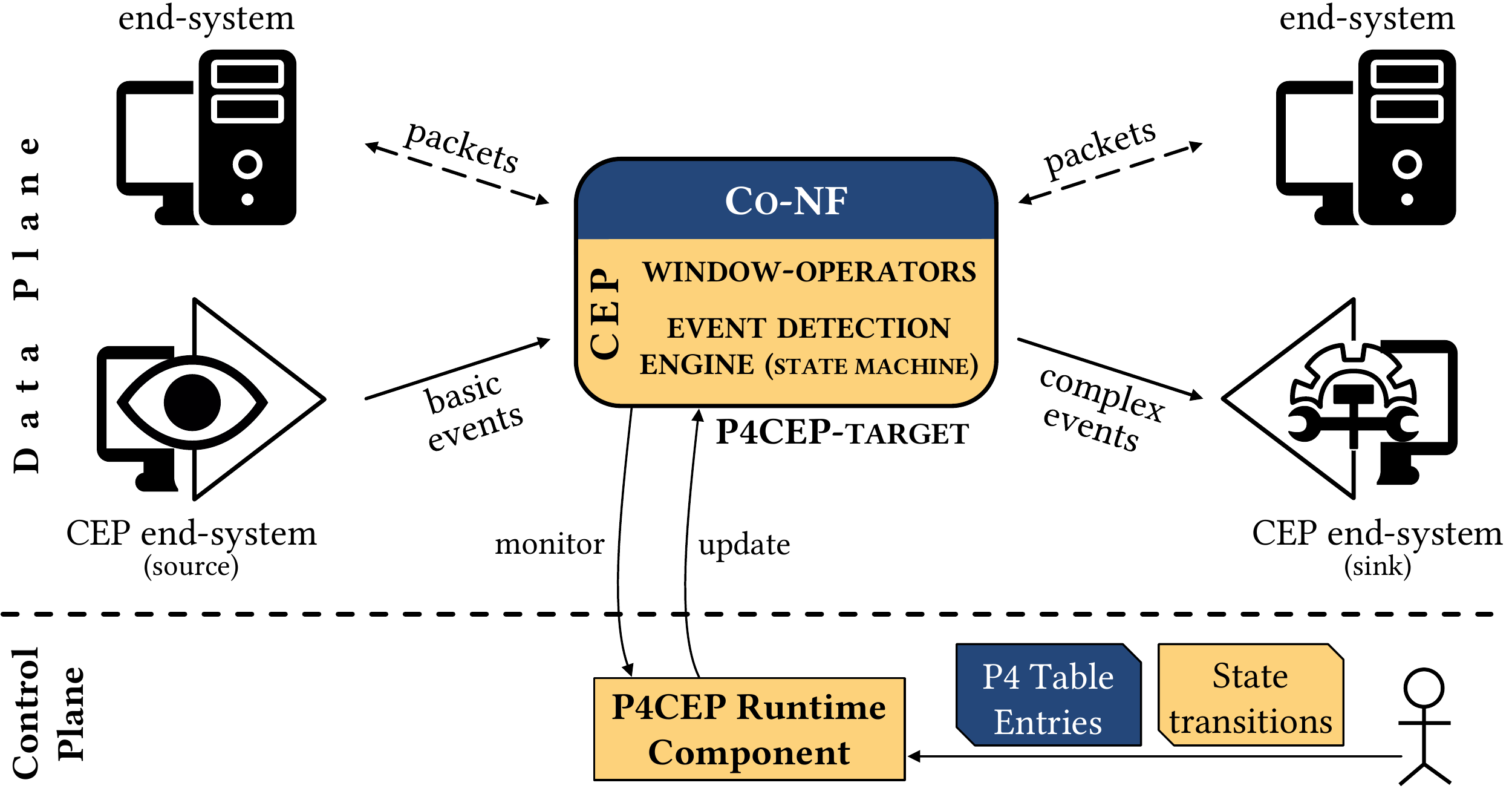}
	\caption{The P4CEP system model.}
	\label{fig:system-model}
\end{figure}

P4CEP's system model, illustrated in \Cref{fig:system-model}, assumes a set of \textbf{end-systems} that are interconnected by a set of programmable network processing elements (\textbf{P4CEP-targets}), forming a data plane topology.
End-systems that host event-based applications (CEP end-systems) are differentiated into \textbf{event sources}, which observe \textbf{basic events} and disseminate them, e.g., networked-sensors or server reporting performance metrics or log data, and \textbf{event sinks}, which receive and react to \textbf{complex events}.
P4CEP-targets (listed in §\ref{subsec:background:P4}) implement two types of functions: (1)~network~functions, which co-exist with CEP (\textbf{Co-NF}, dark-shaded), typically  simple forwarding of non-CEP packets, and (2)~CEP~functions (light-shaded), which can be divided into \textbf{window operators}, which store the $n$ last values of header fields in a FIFO-manner and offer aggregation functions over these values, and the \textbf{event detection engine}, which detects complex events based on a state machine implementation.
Note that without loss of generality we henceforth consider just a single P4CEP-target and discuss the distribution of CEP onto multiple targets in the roadmap (§\ref{sec:roadmap}).
The \textbf{P4CEP runtime component} implements a control plane interface for an operator to P4CEP-targets. Besides deploying compiled P4CEP programs, it handles all runtime tasks: updating P4 table entries and state transitions in the CEP engine as well as acquiring statistics and other monitoring data from the targets.

\subsection{The P4CEP Workflow and Compiler}
\label{subsec:P4CEP:workflow-compiler}
\begin{figure}
	\centering
		\includegraphics[width=\columnwidth]{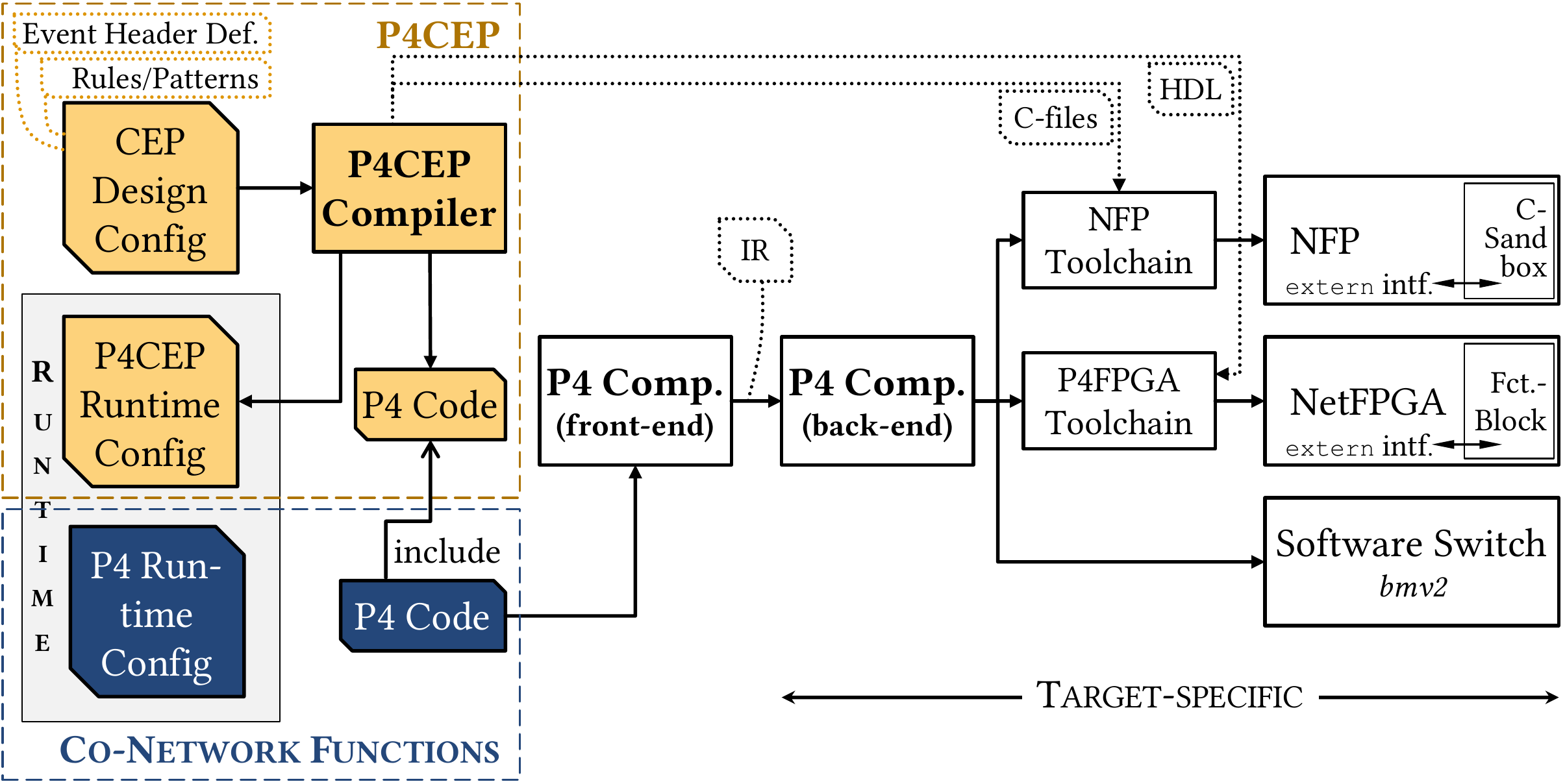}
	\caption{P4CEP workflow: design-time components and source files involved in building P4CEP for different targets.}
	\label{fig:design-workflow}
\end{figure}

P4CEP's design-workflow is illustrated in \Cref{fig:design-workflow}. It is mainly composed of our P4CEP compiler and an unmodified \textbf{P4 compiler} chain, consisting of a target-independent and target-dependent compiler, as well as \textbf{target-dependent toolchains}. Currently, we support target-specific external functions for the Netronome NFP target.
The user-input to the P4CEP compiler (\textbf{CEP design config}) consists of P4-definitions of header fields and parser instructions for packets that are to be interpreted and processed as basic events as well as declarations of window operators and event definition rules, which describe how complex events are derived from basic events. We introduce our CEP rule specification language later (§\ref{subsec:P4CEP:language}).

From these definitions, the \textbf{P4CEP compiler} creates corresponding P4 source code, supporting both P4 versions ($\text{P4}_\text{14}$ and $\text{P4}_\text{16}$). It comprises definitions of registers (inter-packet state, henceforth called global state), metadata structures (per-packet state), auxiliary tables for (multiple) windows, implemented as register ring-buffers, and (multiple) state machines, each associated with a distinct complex event pattern to detect.
When required, our implementation ensures consistency of global data to avoid inconsistency effects such as lost updates, which would translate to unprocessed events or partially, i.e., non-atomically processed events.
We protect global state from concurrent access by implementing critical sections in atomic control flow blocks in $\text{P4}_\text{16}$ and using NFP's atomic register access or it's synchronization library, as described in section \ref{subsec:background:P4}. This inflicts execution overhead in terms of additional latency (\textbf{\emph{Limitation 1}}).

The main logic is implemented in the CEP ingress \textbf{control flow}, which upon receiving a packet that transports basic events executes a sequence of actions implementing \textbf{window operations}, followed by state machine executions.
Due to space constraints, we provide only a high-level description. We make a complete and annotated control flow example available in our P4CEP release \cite{p4cep-release}. First, the current instance count (ringbuffer head pointer) of the window is read from a register and incremented with overflow handling. One drawback of P4 (\textbf{\emph{Limitation 2}}) is that registers cannot be directly referenced in arithmetic operations or as table keys. Thus, register values have to be copied into dedicated intermediate metadata fields and back, which bloats code space and execution overhead.
Then, the header field value of the current event and instance count are stored in registers, i.e., are persisted in the window. For applying the aggregation function on the window, our compiler has to unroll the window iteration, due to P4's lack of loops (\textbf{\emph{Limitation 3}}). The aggregate value is stored in a metadata field $m_{\textit{aggr}}$, as is the iteration counter $m_{\textit{iter}}$.
For each value $r_i$ in the window, $r_i$ has to be copied from the window register to a metadata field $m_i$. Then, the aggregation function 
is applied on $m_{\textit{aggr}}$, referencing $m_i$.
This sequence is repeated for all windows.

A pattern of basic events defines a complex event whose detection is modeled as a deterministic finite \textbf{state machine} ${C=(\Sigma, S, s_0, \delta, F)}$, as illustrated in \Cref{fig:p4cep-state-machine}.
It consists of a sequence of basic events (input symbols $x \in \Sigma$), where typically a basic event is specified by \textbf{predicates} $P_x$ (simple or compound) on packet header fields, as we describe in greater detail in §\ref{subsec:P4CEP:language}. For each pattern, the following actions are executed sequentially: all packet predicates are evaluated.
If $P_x$ evaluates to true, an id associated to that predicate ($P_x\! \rightarrow \! x$) is stored in a metadata field $m_x$. Then, the state machine is executed by first acquiring the current state $q\! \in\! S$ by copying from a register to a metadata field $m_q$, followed by performing a lookup with the key-pair <$m_q, m_x$> on the \textbf{transition table}---a P4 table encoding $\delta$. Upon a match, the returned value pair <$ {q_n\!=\!\delta(q,x)}, \  b_{\textit{is\_accepting}}$> is written to registers if $q_n$ is not an accepting state ($\neg b_{\textit{is\_accepting}} \equiv q_n \notin F$). If it is an accepting state, the state machine is reset, i.e., $q_n$ is set to the initial state $s_0$, and the return value for the complex event is set, encoded in a header field, before the packet is sent to registered CEP sinks using the P4-\texttt{resubmission} mechanism.
P4CEP allows the detection of multiple complex events by sequential execution of the corresponding state machines.

The P4 code generated by the P4CEP compiler is merged with the user-provided P4 program source file, which implements co-NF functionality, using P4's \texttt{include}-primitive. Additionally, runtime configuration files hold table entries and can be \mbox{(re-)deployed} at runtime by the P4CEP runtime control plane component. They are created in a target-compatible format by the P4CEP compiler and given as user-input for the co-NF part, respectively.

\subsection{Limitations for Stateful Processing}
\label{subsec:P4CEP:limitations}
Here, we discuss the encountered limitations of P4 and their implications for stateful packet processing.
Additional to the aforementioned Limitation 1 (synchronizing access to global state), Limitation 2 (no direct operations on registers), and Limitation 3 (lack of a loop construct), another limitation lays in the fact that conditions in P4 can be \emph{only} used within the control flow, not within actions (Limitation 4). Furthermore, $\text{P4}_\text{14}$-actions cannot be directly executed within a control flow, but have to be indirectly executed by using P4's \texttt{apply}-primitive to perform a lookup on an empty dummy-table where the action to be executed is specified as the default action (Limitation 5).
We realize that some of these limitations are inherent design trade-offs in creating P4, which seemed to be driven by satisfying the intricate requirements of switch hardware architectures \cite{jose_compiling_2015} to maintain line-rate processing, for instance ruling out loops (Limitation 3), rather than having stateful packet processing in mind. However, we observe that the evolution of P4 with $\text{P4}_\text{16}$ facilitates stateful packet processing, e.g., by the introduction of the atomicity primitive (Limitation 1) and corrects other seemingly unnecessary limitations like the action indirection (Limitation 5).

\vspace{-2mm}

\subsection{P4CEP Rule Specification Language}
\label{subsec:P4CEP:language}
The specification of the CEP-functionality (Listing \ref{lst:p4cep-example-rules}) is split into the definition of window operators and event definition rules, which are compiled to a corresponding state-machine (\Cref{fig:p4cep-state-machine}).
\begin{figure}[b]
\vspace{-4mm}
\lstset{
	basicstyle=\footnotesize\ttfamily,
  keywordstyle=\bfseries,
  morekeywords={window,size,complex_event,return_value,value,strategy,pattern},
  label={lst:p4cep-example-rules},
  caption={Exemplary P4CEP-rule definition of a window and a sequential pattern, composed of predicates on simple L3/L4-packets and on the window.},
  frame = tb
}
\begin{lstlisting}
window sample_wnd {
  size 8
  value ipv4.totalLen
}
complex_event sample_evt {
  value sum(ipv4.totalLen)
  strategy skip-till-next-match
  pattern ([ipv4.totalLen > 500] && [tcp.dstPort == 80]) ;
          ([sum(sample_wnd) > 6000] ||
           [ipv4.protocol == 17])
}
\end{lstlisting}
\end{figure}
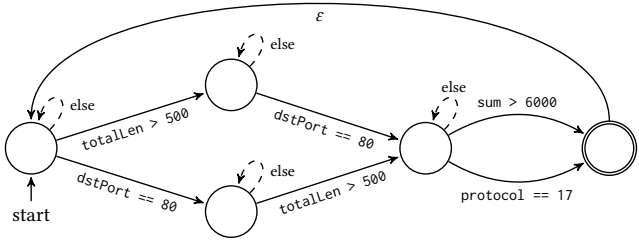
\begin{figure}[t]
	\centering
\resizebox{\columnwidth}{!}{%
\begin{tikzpicture}[->,>=stealth',shorten >=1pt,auto,semithick]
   \tikzstyle{every state}=[draw]
   \node[initial below,state] (A)                    {};
   \node[state]         (B) [above right=0.4cm and 2.5cm of A] {};
   \node[state]         (D) [below right=0.4cm and 2.5cm of A] {};
   \node[state]         (C) [right=5.25cm of A] {};
   \node[state, accepting] (E) [right=2cm of C] {};
 
   \path (A) edge              node [below, sloped] {{\footnotesize \texttt{totalLen > 500}}} (B)
             edge              node [below, sloped] {{\footnotesize \texttt{dstPort == 80}}} (D)
						 edge [in=75, out=45, loop, dashed] node [pos=0.35, right] {\footnotesize else} (A)
         (B) edge              node [below, sloped] {{\footnotesize \texttt{dstPort == 80}}} (C)
				     edge [in=75, out=45, loop, dashed] node [pos=0.3, right] {\footnotesize else} (B)
         (C) edge [bend left]  node [] {{\footnotesize \texttt{sum > 6000}}} (E)
             edge [bend right] node [below] {\footnotesize \texttt{protocol == 17}} (E)
						 edge [in=75, out=45, loop, dashed] node [pos=0.5, above] {\footnotesize else} (C) 
         (D) edge              node [below, sloped]{\footnotesize \texttt{totalLen > 500}} (C)
				     edge [in=75, out=45, loop, dashed] node [pos=0.3, right] {\footnotesize else} (D) 
				(E) edge [looseness=0.695, in=200, bend right=90] node {$\varepsilon$} (A); 
\end{tikzpicture}
}
	\caption{Generated finite state machine detecting complex event patterns as specified in Listing \ref{lst:p4cep-example-rules}.}
	\label{fig:p4cep-state-machine}
\end{figure}
Note that for sake of illustration, we simplified the example by using headers of common layer 3 and 4 protocols. Instead, custom headers possibly encoding type and value of basic events could be used.
Moreover, we used simple instead of composed predicates.

As a side effect, the example shows that some co-NFs, such as anomaly/intrusion-detection, typically relying on sums or counts of specific (sequence of) packets, can be mapped directly to CEP-functionality. The illustrated example (Listing \ref{lst:p4cep-example-rules}, \Cref{fig:p4cep-state-machine}) enables the detection of the following anomaly pattern: a large IPv4 packet and an HTTP-packet, followed in sequence by an UDP-datagram or the sum of total lengths over all last $n=8$ seen IPv4 packets exceeding 6 KB.

The following concepts of the P4CEP rule specification language are used to express such patterns:

\textbf{\texttt{Window} definitions} consist of the window \textbf{\texttt{size}} $n$ and a field reference whose \textbf{\texttt{value}} is to be stored within that window (last eight IPv4 total lengths in the example). Field references are simple references to P4 headers or metadata that must have been parsed by the P4 program and thus be defined either as a CEP event header or as a co-NF header. Windows can be referenced by name within a pattern of a complex event definition or as its return value.

The definition of \textbf{\texttt{complex\_event}s} is structured as follows:
(1)~A return \textbf{\texttt{value}} to be set in a complex event packet, sent in case of detection. This can be (1a)~any valid P4 expression (static expression, field reference), or~(1b) a reference to an aggregation function over a window (e.g., \texttt{sum(sample\_wnd)}), or over a header field (see example).
\newline\noindent
(2)~A \textbf{\texttt{strategy}} specifying the state transition if an incoming basic event does not match any predicate:
\newline\noindent
(2a)~\texttt{skip-till-next-match} performs a transition to the same state, i.e., ignores the event, (\texttt{else}-branches in \Cref{fig:p4cep-state-machine} of the example);
(2b)~\texttt{strict} resets the state-machine by setting the next state to the initial state.
\newline\noindent
(3)~A \textbf{\texttt{pattern}} of basic events defining a complex event. Basic events are specified by simple or compound predicates. A predicate can be any valid P4 condition on one or more field references, or a condition on an aggregation function over a window or over a header field.
Predicates are demarcated by square brackets and combined to patterns using the following logical operators.
(3a)~Sequence~\texttt{;}: the left predicate must hold true before the right.
(3b)~Conjunction~\texttt{\&\&}: both predicates must hold true (in any order).
(3c)~Disjunction~\texttt{||}: one of the predicates must hold true.

Finally, the following \textbf{aggregation functions} on windows or field references are currently supported:
\texttt{sum}, \texttt{min}, \texttt{max}, and \texttt{count} (which counts how many times a predicate was true).
We plan to implement \texttt{average}, which is not straightforward due to P4's missing float support and lack of a division operator, but can be approximated by fix-point and bit-shift operations on windows of sizes $2^n$.
%
%
%
\section{Preliminary Evaluation}
\label{sec:eval}
In this section, we provide a first impression of P4CEP's practicability, evaluated on state-of-the-art P4 targets:
a Netronome Agilio smart-NIC with two 10GbE-ports, on which we run a pure P4 implementation of P4CEP (\textbf{NFP}) and an optimized version employing \texttt{extern}-functionality through NFP's C-sandbox (\textbf{NFP-C}), as well as the software-switch \textbf{bmv2}.

We first provide a baseline analysis of a simple P4 program, implementing stateless forwarding based on parsing layer 2--5 headers of smallest-sized packets. We measured a baseline latency including serialization delay of $\SI{6.8}{\micro\second}$ for NFP(-C) and $\SI{475}{\micro\second}$ for bmv2, respectively. Baseline throughput is full line-rate ($\approx$14.88 million packets per second (Mpps) for 10GbE) for NFP(-C) and 0.08\% thereof ($\approx$ 12 Kpps) for bmv2, denoted as relative throughput $B_p$.

\begin{figure}[t]
	\centering
		\includegraphics[width=\columnwidth]{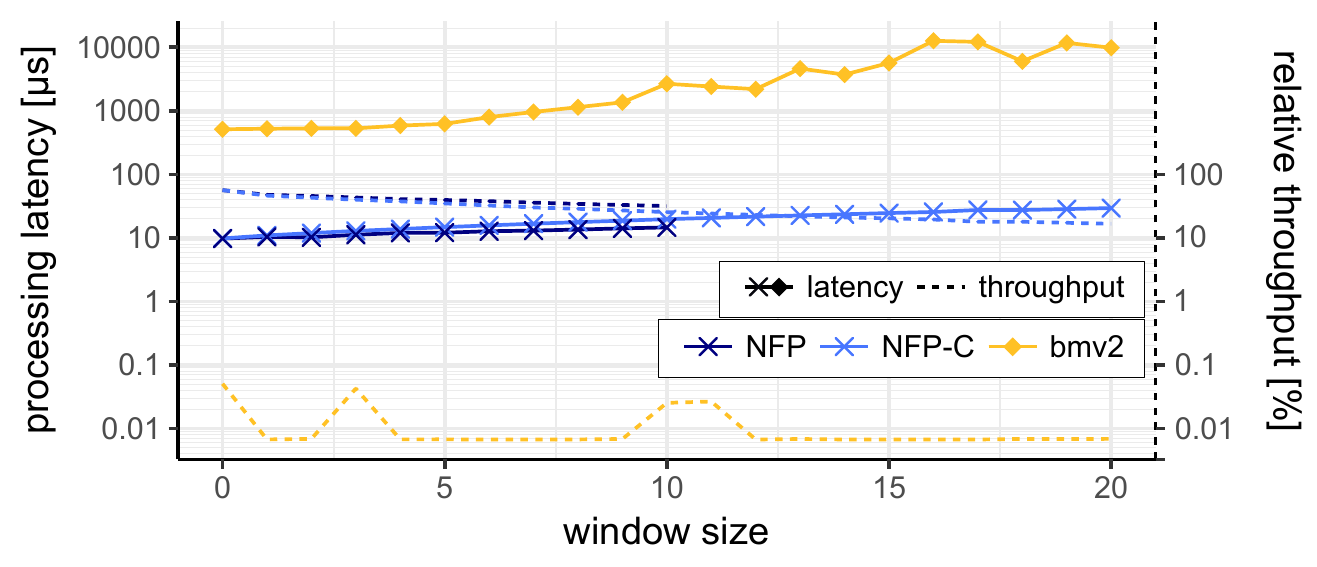}
	\caption{P4CEP's performance for increasing window sizes on an NFP smart-NIC and bmv2.}
	\label{fig:eval}
\end{figure}

We approximate the latency for CEP processing $l_p$ by hardware-timestamping the egress of basic events, sent at a rate of 500 pps, and the ingress of consequently detected complex events. Propagation and serialization delay are negligible. \Cref{fig:eval} shows the mean performance over 60,000 samples for varying window sizes $n$ and one complex event pattern to be detected consisting of two basic events with one simple predicate each. In the depicted interval ${0\! \leq\! n\!\leq\! 20}$, we measured a processing latency of \mbox{\SI{9.8}{\micro\second} $\leq\! l_p\! \leq$ \SI{29.5}{\micro\second}} and relative throughput of $56\% \geq\! B_p\! \geq 16\%$ for NFP-C. The pure P4 implementation (NFP) performs slightly better, showing low overhead for the \texttt{extern}-mechanism. However, for $n>10$, the generated P4 code exceeds the size limit of NFP and is hence rejected. With window handling implemented in the C-sandbox, NFP-C scales up linearly with $\Delta l_p\approx$~\SI{1}{\micro\second} per iteration, up to $l_p\approx$~\SI{969}{\micro\second} for $n=1000$, where $B_p$ drops to 0.4\%~($\approx$60 Kpps). Overall, NFP(-C)'s standard deviation of $l_p$ (jitter) is quite low (tens to hundreds of nanoseconds), as is its deviation of throughput ($\approx$0.02\%).

While bmv2 has no code size restrictions, its performance is significantly worse. Starting with $l_p \approx$ \SI{512}{\micro\second} and $B_p \approx$~0.05\%, 
it also shows inferior scalability properties, exceeding  $l_p \approx$ \SI{10}{\milli\second} for $n\! >\! 15$. Jitter is also increased.

We conclude that already in its early stage, P4CEP achieves good performance in particular on hardware targets. We expect even higher performance on more optimized implementations and targets (NetFPGA, PISCES) in the future.
%
%
%
%
\section{Related Work}
\label{sec:related-work}
In this section, we briefly discuss related work other than already mentioned.

SNAP \cite{arashloo_snap:_2016} is a network-centric, high-level language for network programming, extending stateless packet processing with primitive stateful operations.
It offers a stateful network-wise abstraction for packet processing by enabling access to a persistent global array in control programs, while making the distribution of that state in the data plane transparent to the programmer. Packet processing in a SNAP program depends on the current state of the network, which is held in variables within the global array and which is possibly changed as a result of processing. SNAP considers events as non-frequent changes in the network, such as traffic changes and failures, that trigger recompilation of the network program in the control plane.
 
Stateful NetKAT is a language for event-driven network programming \cite{mcclurg_event-driven_2016}, extending the NetKAT language by mutable state. It is similar to SNAP, however, it rather focuses on applying consistent network updates where consistency properties hold during the transition between two network configurations, which is triggered in response to events.

While both approaches enable network-centric stateful packet processing, P4CEP is tailored for complex event processing with a generic notion of events that includes but is not limited to network events. Since P4CEP is based on P4, it is more lightweight while still leveraging the full expressiveness of P4.

OpenState \cite{bianchi_openstate:_2014} is an implementation of a generic state machine in the data plane of an OpenFlow switch. It maps the state machine execution to a fixed match-action pipeline consisting of two tables, holding the current state and transitions on a per-flow basis. Since state is held in flow table entries, OpenState requires a custom OpenFlow instruction to be able to update the state after a transition. While the authors provide a modified implementation of an OpenFlow software switch, there are no hardware implementations. P4CEP uses P4 to implement its state machine logic without the need for modifications of software or hardware.
%
%
%
\section{Conclusion \& Roadmap to a Distributed In-Network CEP}
\label{sec:roadmap}
In this paper, we presented P4CEP, an in-network implementation of Complex Event Processing, and showed its practicability by experiment.

Based on our experiences gained from designing P4CEP, we argue that in-network computation, in particular for stateful processing, poses an interesting research question regarding the trade-off between portability (target-independence) and leveraging programmability, including application-specific custom functions (introducing target-dependence).
For instance, while it was our design goal to stay target-independent through exclusive use of a uniform data-plane programming language (P4), implementing custom functions enabled mitigation of current limitations of P4 and enriched functionality at the cost of becoming target-dependent.
In summary, we identified the following limitations of P4 for stateful in-network computing: (1) the overhead of state synchronization, (2) the inability to directly handling global state in registers, and (3) the indirection of action invocations. Although we understand some limitations as deliberate decisions in P4's design, we see great potential for constructs like bounded loops or more efficient primitives for synchronization.
To further explore this trade-off, we plan to adopt more powerful operators from CEP and stream processing.

Another means to counter the observed limitations and to increase resource-efficiency is to leverage the distribution of in-network computation.
For CEP, distribution bears the following benefits:
(1)~The early filtering of needless basic events, i.e., events that are not part of any complex event pattern, leads to reduced load in the event detection engine and to increased bandwidth-efficiency. Thus, we plan to adopt our in-network content-based publish/subscribe approach \cite{bhowmik_high_2017} using data plane programming to increase the expressiveness of filtering.
(2)~The disaggregation of event detection to multiple P4CEP-targets allows for the parallelization of CEP and hence increases performance. This requires strategies for replication and partitioning of basic events and a concept of splitting window handling and event detection, which we plan to develop. Ideally, distribution compensates some of the observed target limitations, such as code-size and pipeline-depth limits.

\balance
\bibliographystyle{ACM-Reference-Format}
\bibliography{reference}

\end{document}